\documentclass[12pt]{article}
\usepackage{psfig}
\begin{document}

\title{Phase coherent quantum mechanical spin transport \\
in a weakly disordered quasi one-dimensional channel}
\author{M. Cahay $\thanks{Corresponding author. E-mail: marc.cahay@uc.edu}$\\
Department of Electrical and Computer Engineering and Computer Science\\
University of Cincinnati, Cincinnati, Ohio 45221\\
\\S. Bandyopadhyay\\
Department of Electrical Engineering\\
Virginia Commonwealth University\\
Richmond, Virginia 23284}

\date{}

\maketitle

\baselineskip=24pt

\begin{center}
{\bf Abstract}
\end{center}

\medskip

\noindent
A transfer matrix technique is used to model phase coherent spin transport in 
the weakly disordered quasi one-dimensional channel of a gate-controlled 
electron spin interferometer [Datta and Das, Appl. Phys. Lett., \underline{56}, 
665 (1990)].  
%The model is fully quantum mechanical and goes beyond the 
%classical drift-diffusion or Boltzmann transport type models that have been 
%typically employed in the past to model spin transport. 
It includes the effects of an axial magnetic field in the 
channel  of the interferometer (caused by the ferromagnetic contacts), a Rashba 
spin-orbit interaction, and elastic (non-magnetic) impurity scattering. We show 
that in the presence of 
the axial magnetic field, non-magnetic impurities can cause spin relaxation in a 
manner similar to the Elliott-Yafet mechanism.  The amplitudes and phases of 
the conductance oscillations of the interferometer, and  the degree of 
spin-conductance polarization, 
are found to be quite sensitive to the height of the interface barrier at the 
contact, as well as  the strength, locations and nature (attractive or 
repulsive) of just a few elastic non-magnetic 
impurities in the channel. This can seriously hinder practical applications of 
spin interferometers.

\bigskip

\noindent {\bf PACS}: 72.25.Dc, 72.25.Mk, 73.21.Hb, 85.35.Ds

\pagebreak

\section{Introduction} 

In a seminal paper published in 1990, Datta and Das \cite{datta} proposed a  
gate controlled electron spin interferometer which is an analog of the standard 
electro-optic light modulator. Their device consists of a  one-dimensional 
semiconductor channel with ferromagnetic source and drain contacts (Fig. 1). 
Electrons are 
injected  into the channel from the ferromagnetic source with a definite spin, 
which is 
then controllably precessed in the channel with a gate-controlled Rashba 
interaction  \cite{rashba}, 
and finally  sensed 
at the drain. At the drain end, the electron's transmission probability depends 
on the relative alignment of its spin with  the drain's (fixed) magnetization.  
By controlling 
the angle of spin precession in the channel with a gate voltage, 
one can modulate the relative spin alignment at the drain end, and hence control 
the source-to-drain  current (or conductance). In this device, the ferromagnetic 
contacts act as ``spin polarizer'' (source) and ``spin analyzer'' (drain).

%NEW STUFF
There have been some studies of ballistic spin transport in such a device 
\cite{mireles1,mireles2,add2,add3}, 
but they did not consider two features that are always present in a real 
device structure. 
%END OF NEW STUFF
First, there is an axial magnetic field along the channel 
caused by the ferromagnetic contacts. This field dramatically alters the 
dispersion relations of the subbands in the channel, causes spin mixing, and has 
a serious 
effect on spin transport.  Second, there will always be a few impurities in the 
channel 
(even if they are remote impurities) associated with channel doping. We show 
that these impurities, even if they are non-magnetic, can cause spin relaxation 
in the presence of the axial magnetic field. Thus, they can affect the 
conductance modulation of the interferometer and the degree of spin polarization 
of the current.

This paper is organized as follows. In the next section, we describe the 
Hamiltonian to model the gate-controlled electron spin interferometer depicted 
in Fig. 1.
The Hamiltonian includes  potential barriers at the contact/channel interface 
that are inevitably present, the axial magnetic field, and localized impurities 
in the channel. It does not include 
perturbations due to phonons and other time dependent scattering potentials (we 
assume that 
the channel is shorter than the phase breaking length so that transport is phase 
coherent). Using a truncated form of this Hamiltonian, we derive the dispersion 
relations of the subbands 
in the channel. Because of the presence of the axial magnetic field, the 
subbands are not 
eigenstates of the spin operator. Therefore, no subband has a definite spin 
quantization axis. 
Furthermore, eigenspinors in two subbands (at the same energy) are {\it not} 
orthogonal. As a result,  elastic 
(non-magnetic/spin-independent) impurity scattering can couple two subband 
states with non-orthogonal eigenspinors, 
causing elastic inter-subband transitions that {\it relax spin}.  One should  
compare this 
mechanism of spin relaxation with the Elliott-Yafet  spin relaxation mechanism 
\cite{elliot}  
in a bulk semiconductor. The Elliott-Yafet  relaxation comes about because in a 
real crystal, the Bloch 
states are not eigenstates of spin so that an ``upspin'' state has some 
``downspin'' component and vice versa. As a result, non-magnetic impurity 
scattering can connect (mostly) upspin and (mostly) downspin electrons leading 
to a spin relaxation. Our mechanism is very similar.

Section III contains numerical examples of the conductance modulation
of a spin interferometer as a function of  applied gate potential, spin 
polarization of the current through the channel, and effects of the interface 
barriers and elastic 
(non-magnetic) impurity scattering. Finally, section IV contains our 
conclusions. 

\newpage
\section{Theoretical model}

We first consider the quasi one-dimensional semiconductor channel of a spin 
interferometer in the absence of any impurities.  The channel is along the 
x-axis (Fig. 1) and the gate electric field is 
applied along the y-direction to induce a Rashba spin-orbit coupling in the 
channel. This 
system  is described by the single particle effective-mass Hamiltonian 
\cite{moroz}
\begin{equation}
{\cal H} = {{1}\over{2 m^*}} \left ( {\vec p} + e{\vec A} \right )^2 
+ V_I (x)
+ V_1 (y) + V_2 (z) 
-  (g^{*}/2) \mu_B {\vec B} \cdot {\vec \sigma} 
+  \frac{{\alpha}_R}{ \hbar} \hat{y}\cdot \left [ {\vec \sigma} \times ( {\vec 
p} + e {\vec A} ) 
\right ]
\label{hamiltonian}
\end{equation}
where $\hat{y}$ is the unit vector along the y-direction in 
Fig. 1 and ${\vec A}$ is the vector potential due to the axial magnetic  
field ${\vec B}$ along the 
channel (x-direction) caused by the ferromagnetic contacts. In Equation (1), $ 
\mu_B $ 
is the Bohr
magneton ( $ e \hbar / 2 m_0$) and $g^* $ is the effective Land\'e g-factor of 
the 
electron in the channel.  
The quantity 
$\alpha_R$ is the Rashba spin-orbit coupling strength which  can be varied with 
the gate potential.
The confining potentials along the y- and z-directions are denoted by $V_1 (y)$ 
and 
$V_2 (z)$, with the latter being parabolic in space.

In Equation (1), $V_I(x)$ represents an interfacial potential barrier 
between the ferromagnetic contacts and the semiconducting channel. If the 
contact neighborhood consists of heavily doped semiconductor material in close 
proximity to a metallic ferromagnet, 
the Schottky 
barriers at the interface will be very narrow \cite{aschwalom} and electrons 
from the contacts can tunnel fairly easily into semiconducting channel resulting 
in a nearly-ohmic contact.
We model these ultra-narrow Schottky barriers as delta-barriers given 
by:
\begin{equation}
V_I(x) = V_L \delta (x) + V_R \delta (x-L) 
\end{equation}
where $ V_L $ and $V_R $ are assumed equal. In practice, the strength of 
the barrier depends
on the ferromagnetic materials and also on the doping level in the channel.
These barriers have a beneficial effect; they can facilitate coherent spin 
injection 
across a metallic ferromagnet and a semiconducting paramagnet interface 
\cite{eirashba} which is crucial for a spin interferometer.

In Equation (\ref{hamiltonian}), we have neglected a few effects for the sake of 
simplicity. We have neglected the normal Elliott-Yafet interaction \cite{elliot} 
because it is weak in quasi one-dimensional structures (where elastic scattering 
is strongly suppressed \cite{sakaki}). We have also neglected  the Dresselhaus 
interaction \cite{dresselhaus} since it does not relax spin when the initial 
spin polarization is along the axis of the wire 
\cite{bournel,privman1,privman2,pramanik} (this is 
the case with the 
gate controlled spin interferometer). The Dresselhaus interaction can however be 
easily included in 
the Hamiltonian  and is left for future work. Finally, we model localized 
non-magnetic 
impurities (i.e., which
do not flip the spin) using a standard model of 
delta-scatterers. The scattering potential is given by 
\begin{equation}
V_{imp} = \sum_{i=1}^{N} {\Gamma}_i \delta ( x - x_i)
\end{equation}
to represent N impurities in the channel at location $x_i$ and with strength 
${\Gamma}_i$ (assumed to be spin independent). In our numerical
examples, we consider the case of both attractive (${\Gamma}_i$ negative) and 
repulsive (${\Gamma}_i$ positive) impurities. While Equation (1) represents a 
ballistic channel with no scattering, addition of the scattering potential in 
Equation (3) to Equation (1) will result in a Hamiltonian describing a weakly 
disordered channel in which impurity scattering takes place. The eigenstates of 
this (spin-dependent) Hamiltonian can then be found using a transfer matrix 
technique to extract the electron wavefunction in the presence of impurity 
scatterers. From this wavefunction, we can calculate the (spin-dependent) 
transmission probability through the channel and ultimately the (spin-dependent) 
channel conductance.

The choice of the Landau gauge ${\vec A}$ = (0, -Bz, 0)  allows us to decouple 
the y-component of the Hamiltonian in Equation (2) from the x-z component.
Furthermore, if we ignore $V_I(x)$ and $V_{imp}$ which are delta potentials, the 
rest of the  Hamiltonian is translationally invariant in the 
x-direction. Therefore, the wavevector $k_x$ is a good quantum number in a 
ballistic channel and the eigenstates 
are plane waves traveling in the x-direction.  The two-dimensional Hamiltonian 
in the plane of such a channel 
(x-z plane) is then given by
\begin{equation}
H_{xz} = {{p_z^2}\over{2 m^*}} + \Delta E_c +  {{1}\over{2}}m^* \left ( 
\omega_0^2 + 
\omega_c^2 \right ) z^2 + {{\hbar^2 k_x^2}\over{2 m^*}} +{{\hbar^2 k_R 
k_x}\over{m^*}} \sigma_z - ( g^* /2) \mu_B B {\sigma}_x  - {{\hbar k_R 
p_z}\over{m^*}} \sigma_x
\label{Hamiltonian}
\end{equation}
where $\omega_0$ is the curvature of the confining potential in the z-direction,
$\omega_c$ = $eB/m^*$,  $k_R = m^* 
\alpha_R/\hbar^2$, and $\Delta E_c$ is the potential barrier between the 
ferromagnet and 
semiconductor. We assume that $\Delta E_c$ includes the effects of the quantum 
confinement in the y-direction. 

The scattering potential $V_{imp}$ and the interface potential $V_I(x)$ couple 
various wavevector states $k_x$. This is handled by the transfer matrix 
technique described later.

\subsection{Energy dispersion relations}
 
We now derive the energy dispersion relations in the channel of a ballistic 
interferometer using Equation (\ref{Hamiltonian}).  The first five
terms of the Hamiltonian in Equation (\ref{Hamiltonian}) 
yield shifted parabolic subbands with dispersion relations:
\begin{equation}
E_{n, \uparrow} = ( n + 1/2) \hbar \omega + \Delta E_c
+ {{\hbar^2 k_x^2}\over{2 m^*}} + {{\hbar^2 k_R k_x}\over{m^*}}, ~~~
E_{n, \downarrow} = ( n + 1/2) \hbar \omega + \Delta E_c
+ {{\hbar^2 k_x^2}\over{2 m^*}} - {{\hbar^2 k_R k_x}\over{m^*}},
\label{dispersion}
\end{equation}
where $\omega = 
\sqrt{ \omega_0^2 + \omega_c^2 }$. In Equation (\ref{dispersion}), the 
$\uparrow$ 
and $\downarrow$ 
arrows indicate +z and -z 
polarized spins (eigenstates of the $\sigma_z$ operator) which are split by the 
Rashba effect (fifth term of the Hamiltonian in Equation (\ref{Hamiltonian}). 
These are 
subbands with definite 
spin quantization axes along +z and -z directions since they are eigenstates of 
the $\sigma_z$ operator. Their dispersion relations 
are shown as dashed lines in Fig. 1.

The sixth and seventh  terms in Equation (\ref{dispersion}) induce a 
perturbation 
and mixing between the unperturbed subbands (+z- and -z-polarized spins). 
The sixth term originates from the magnetic field due to the ferromagnetic 
contacts and the seventh originates from the Rashba effect itself.   The ratio 
of 
these two terms can be shown to be 
of the order of 10$^4$ - 10$^6$ for typical values of the relevant parameters. 
Therefore, we can neglect the seventh term in comparison with the sixth term.

To obtain an analytical expression for the dispersion relation 
corresponding to the first six terms in the Hamiltonian in 
Equation (\ref{Hamiltonian}),  we derive the two-band dispersion relation in a 
truncated Hilbert space considering 
mixing between the two lowest unperturbed subband states (namely the +z and -z 
spin states).
Straightforward diagonalization of the Hamiltonian in Equation 
(\ref{Hamiltonian}) 
(minus the seventh term) in the basis of these two unperturbed states gives the 
following 
dispersion relations:
\begin{equation}
E_1 (k_x) = {{1}\over{2}} \hbar \omega + \Delta E_c + {{\hbar^2 k_x^2}\over{2 
m^*}}
- \sqrt{ \left ({{\hbar^2 k_R k_x}\over{m^*}} \right )^2 + \left (
\frac{ g^* \mu_B B}{2} \right )^2 },
\label{dispersion1}
\end{equation}
\begin{equation}
E_2 (k_x) = {{1}\over{2}} \hbar \omega + \Delta E_c + {{\hbar^2 k_x^2}\over{2 
m^*}}
+ \sqrt{ \left ( {{\hbar^2 k_R k_x}\over{m^*}} \right )^2 + \left (
\frac{ g^* \mu_B B}{2} \right )^2 },
\label{dispersion2}
\end{equation}
where the indices 1 and 2 refer to the lower and upper subbands. Their 
dispersion relations are plotted schematically as solid lines in Fig. 1.

One can see from Fig. 1 that the magnetic field caused by the ferromagnetic 
contacts couples the two unperturbed subbands (the original +z and -z-polarized 
subbands) and changes their dispersion 
relation, lifting the degeneracy at $k_x$ 
= 0. While the unperturbed bands are shifted parabolas with single minima at 
$k_x = \pm k_R$ \cite{datta}, 
the perturbed bands (in the presence of a magnetic field) are not parabolic and 
are symmetric about the energy axis. One of them has a single minimum at $k_x$ = 
0, and the 
other has double minima at $k_x = \pm k_R \sqrt{1 + ( g^* \mu_B B / 
\delta_{R})^2}$,
where $\delta_{R}$ = ${\hbar}^2 k_R^2/2m^*$. The magnetic field not only has 
this profound influence on the dispersion relations, but it also causes {\it 
spin 
mixing}, meaning that the perturbed subbands no longer have definite 
spin quantization axes (they are no longer +z and -z-polarized subbands) because 
they are no longer eigenstates of the spin operator.
Spin quantization becomes wavevector dependent. 
Furthermore, energy-degenerate states in the two perturbed subbands no longer 
have orthogonal spins. Therefore, elastic scattering between them is possible 
without a complete spin flip. 

The energy dispersion relations also show that  the difference $\Delta k_x$ 
between the wavevectors in the 
two subbands at any given energy is {\it not} 
independent of that energy. Since $\Delta k_x$ is proportional to the angle by 
which the spin precesses in the channel \cite{datta}, 
the angle of spin precession in the channel os a spin interferometer is {\it no 
longer} independent of electron energy. 
Thus different 
electrons that are injected from the contact with different energies (at finite 
temperature and bias) will undergo different 
degrees of spin precession, and the conductance modulation will not survive 
ensemble averaging over 
a broad spectrum of electron energy at elevated temperatures and bias. In ref. 
\cite{datta}, which did not consider the effect of the axial magnetic field,
a point was made that the angle of spin precession is independent of electron 
energy so that every electron undergoes the same degree of spin precession in 
the channel
irrespective of its energy. As a result, the conductance modulation of the spin 
interferometer is not diluted by ensemble averaging over electron energy at 
elevated temperature and bias. Indeed this is true in the absence of the 
axial magnetic field, but when the magnetic field is considered, this advantage 
is lost.

From Equations (\ref{dispersion1} - \ref{dispersion2}), we find that an 
electron incident with total energy E has wavevectors in the two channel
subbands given by
\begin{equation}
k_{x \pm} = \frac{1}{\hbar} \sqrt{ 2m^*  ( \frac{ B \pm \sqrt{B^2 - 4C}  }{2}) 
},
\end{equation}
where
\begin{equation}
B = 2 (E - \frac{ \hbar \omega}{2} - \Delta E_c ) + 4 {\delta}_{R}, ~~~
C = ( E - \frac{ \hbar \omega}{2} - \Delta E_c )^2 - {\beta}^2,
\end{equation}
with $\beta$ = $ g^* \mu_B B/2 $.

In Equation (8), the upper and lower signs correspond to the 
lower and upper subbands in Fig. 1 and are referred to hereafter as $k_{x,1}$ 
and 
$k_{x,2}$, respectively.
The corresponding eigenspinors in the two subbands (at energy $E$) are 
respectively
\begin{eqnarray}
\left [ \begin{array}{c}
             C_1 (k_{x,1})\\
             {C^{'}}_1 (k_{x,1})\\
             \end{array}   \right ]
& = &
 \left [ \begin{array}{c}
- \alpha (k_{x,1})/\gamma (k_{x,1})\\
\beta/\gamma (k_{x,1}) \\
\end{array} \right ] \nonumber \\
\left [ \begin{array}{c}
             C_2 (k_{x,2})\\
             {C^{'}}_2 (k_{x,2})\\
             \end{array}   \right ]
& = &
 \left [ \begin{array}{c}
 \beta/\gamma (k_{x,2}) \\
\alpha (k_{x,2}) /\gamma (k_{x,2})\\
\end{array} \right ]
\end{eqnarray}
where the quantities $\alpha$ and $\gamma$ are function of $k_{x}$ and are given 
by
\begin{equation}
\alpha (k_{x})  =  {{\hbar^2 k_R k_x}\over{m^*}} + \sqrt{ \left ( {{\hbar^2 
k_R 
k_x}\over{m^*}} \right )^2 + {\beta}^2 }, ~~~
\gamma ( k_x ) =  \sqrt{ \alpha^2 + \beta^2}.
\label{eigenspinor}
\end{equation} 

Note that the eigenspinors given by Eq. (10) are not 
+z-polarized state $\left 
[ \begin{array}{cc}
1 & 0 
\end{array} \right ]^{\dagger}
$,  or -z-polarized state $\left [ \begin{array}{cc}
0 & 1 
\end{array} \right ]^{\dagger} 
$ if the magnetic field $B \neq 0$. Thus, the magnetic field mixes spins
and the +z or -z polarized states are no longer eigenstates in the channel (in 
other words, the subbands in Eqs. (6) and (7) are not eigenstates of the 
$\sigma_z$ operator unlike the subbands in Equation (5) and hence they are not 
+z and -z-polarized subbands).
Equations (10) also show that the spin quantization (eigenspinor) in any subband 
is not fixed and strongly depends on the wavevector $k_x$. Thus, an electron 
entering the semiconductor channel from the left ferromagnetic contact with 
+x-polarized 
spin, will not couple {\it equally} to +z and -z states. The relative coupling 
will depend on the 
electron's wavevector (or energy). 

Most importantly, the two eigenspinors given by Equation (10) are {\it not} 
orthogonal. Thus, a spin-independent elastic scatterer (non-magnetic impurity) 
can couple these two subbands in the channel and cause elastic inter-subband 
transitions. Another way of 
stating this is that the actual subband states are not eigenstates of the spin 
operator; hence, scattering 
between them is possible via a spin-independent scatterer. This is exactly 
similar to the Elliott-Yafet mechanism in a bulk crystal. Such a scattering 
is of course harmful for the gate controlled spin interferometer since it
introduces a random component to the spin precession in the channel. In our
transfer matrix model (described later) this mechanism of scattering is
automatically included since we use the actual eigenspinors in the channel given 
by Equation (10) to construct the wavefunction (see Section 2.2 later).

We model the ferromagnetic contacts by 
the Stoner-Wohlfarth model. The 
+x-polarized spin (majority carrier) and -x-polarized spin (minority carrier) 
band 
bottoms are offset by an exchange splitting energy $\Delta$ (Fig. 2).

\subsection{Transmission through the interferometer}
 
In this sub-section, we calculate the 
total transmission coefficient through the spin interferometer for an electron 
of energy $E$ 
entering the semiconductor channel from the left ferromagnetic  contact (region 
I) and 
exiting at the right ferromagnetic contact (region III). 
A rigorous treatment of this problem would require an accurate modeling of the 
three- to one-dimensional transition between the bulk ferromagnetic contacts 
(regions I and III) and the quantum wire semiconductor channel (region II) 
\cite{kriman,frohne}. 
However, a one-dimensional transport model to calculate the transmission 
coefficient through the 
structure is known to be a very good approximation when the Fermi wave number in 
the ferromagnetic
contacts is much greater than the inverse of the transverse dimensions of the 
quantum wire \cite{grundler,raichev}. This is always the case with metallic 
contacts.

In the semiconductor channel (region II; $0 < x < L$), the x-component of the 
wavefunction at a position 
$x$ along  the channel is given by
\begin{eqnarray}
{\psi}_{II} (x) & = &
A_I (E)
 \left [ \begin{array}{c}
             C_1(k_{x,1})\\
             C'_1 (k_{x,1})\\
             \end{array}   \right ]
             ^{i k_{x,1} x}
             +
A_{II} (E)
 \left [ \begin{array}{c}
             C_1(-k_{x,1})\\
             C'_1 (-k_{x,1})\\
             \end{array}   \right ]
             e^{-i k_{x,1} x} \nonumber \\
& &   
             +           
A_{III} (E)
 \left [ \begin{array}{c}
             C_2(k_{x,2})\\
             C'_2 (k_{x,2})\\
             \end{array}   \right ]
             e^{i k_{x,2} x}
             +
 A_{IV} (E)
 \left [ \begin{array}{c}
             C_2(-k_{x,2})\\
             C'_2 (-k_{x,2})\\
             \end{array}   \right ]
             e^{-i k_{x,2} x} .
\label{wavefunction}
\end{eqnarray}

For a +x-polarized spin (majority carrier) in the left ferromagnetic contact 
(region I; $x < 0$), 
the electron is spin polarized in the
$\left [ \begin{array}{cc}
             1 1
             \end{array}
             \right ]^{\dagger} $ 
subband  and the x-component of the wavefunction 
is given by
\begin{eqnarray}
{\psi}_I (x) & = &
 \frac{1}{\sqrt{2}} \left [ \begin{array}{c}
             1\\
             1\\
             \end{array}   \right ]
             e^{i {k_{x}}^u x}
             +
  \frac{ R_1 (E)}{\sqrt{2}} \left [ \begin{array}{c}
             1\\
             1\\
             \end{array}   \right ]
             e^{-i {k_{x}}^u x} 
             +
 \frac{R_2 (E)}{\sqrt{2}} \left [ \begin{array}{c}
              1\\
             -1\\
             \end{array}   \right ]
             e^{-i {k_{x}}^d  x}.
\end{eqnarray}
where $R_1(E)$ is the reflection amplitude into the +x-polarized band and 
$R_2(E)$ is the 
reflection amplitude in the -x-polarized band for an electron incident with 
energy $E$.

In the right ferromagnetic contact (region III; $x > L$), the x-component of the 
wavefunction is given by
\begin{equation}
{\psi}_{III} (x)  = 
 \frac{T_1 (E)}{\sqrt{2}} \left [ \begin{array}{c}
             1\\
             1\\
             \end{array}   \right ]
             e^{i {k_{x}}^u  (x-L)}
             +
 \frac{T_2 (E)}{\sqrt{2}} \left [ \begin{array}{c}
             1\\
             -1\\
             \end{array}   \right ]
             e^{i {k_{x}}^d (x-L)}  .
\end{equation}
where $T_1(E)$ and $T_2(E)$ are the transmission amplitudes into the +x and 
-x-polarized bands in the right contact. In Equations (13-14), the wavevectors
\begin{equation}
k_x^u  = \frac{1}{ \hbar} \sqrt{2 m_0 E}, ~~~ k_x^d = \frac{1}{ \hbar} \sqrt{2 
m_0 (E - \Delta)},
\end{equation}
are the x components of the wavevectors corresponding to energy $E$ in the 
majority (+x-polarized) and minority (-x-polarized) spin
bands, respectively.

If there are impurities in the channel,
we must write a solution to the Schr\"odinger equation in each segment of the 
channel between neighboring impurities in the form given by Eq. (12) with 
different
values for the coefficients $ A_i(E) (i=1,4)$.  
In addition to the continuity of the wavefunction across each impurity in the 
channel, the following condition must be satisfied, which is obtained through an 
integration of the Schr\"odinger equation across the impurity:
\begin{equation}
\frac{d \psi }{dx} (x_i + \epsilon )  = \frac{d \psi }{dx} (x_i -\epsilon)  
+ \frac{2 m^* {\Gamma}_i }{ {\hbar}^2 } \psi ( x_i ).
\end{equation}

Furthermore, because of the interfacial barrier at the two 
ferromagnet/semiconductor contacts,
the integration of the Schr\"odinger equation across the left and right 
interface regions lead
to the following two boundary conditions:
\vskip .1in
At x = 0,
\begin{equation}
\mu \frac{d \psi }{dx} (-\epsilon)  + \frac{2 m^* V_0}{ {\hbar}^2 } \psi (0) = 
\frac{d \psi }{dx} (+\epsilon)  +  i k_R (+\epsilon) {\sigma}_z \psi 
(+\epsilon)),
\end{equation}
and, at x = L,
\begin{equation}
\mu \frac{d \psi }{dx} (L+\epsilon)  - \frac{2 m^* V_0}{ {\hbar}^2 } \psi (L) = 
\frac{d \psi }{dx} (L-\epsilon)  +  i k_R (-\epsilon) {\sigma}_z \psi (L),
\end{equation}
where $\mu = \frac{ {m_s}^* }{ {m_f}^* } $ and $ {m_s}^* $ and ${m_f}^*$
are the effective masses in the semiconductor and ferromagnetic materials, 
respectively.
Equations (17) and (18) ensure continuity of the current density at the 
ferromagnetic contact/semiconductor interface.

For the case of two impurities in the channel, the equations above lead to a 
system of 16 equations with 16 unknowns
($R_1(E)$,$R_2(E)$,$T_1(E)$,$T_2(E)$, and three sets of $A_{i}(E)$ 
(i=I,II,III,IV) for the 
three regions in the channel demarcated by
the two impurities). This system of equations must then be solved  to find the 
transmission probabilities $T_1(E)$ and $T_2(E)$. The problem is repeated for
two cases: (i) when the initial spin is +x-polarized (i.e. the incoming 
electron is a majority carrier in the left contact), and (ii) when the incoming 
electron is -x-polarized (i.e. the incident electron is a minority
carrier in the left contact). Finally, the linear response conductance of the 
spin interferometer (for injection 
from either the +x or -x polarized bands in the left contact) is found from the 
Landauer formula

\begin{equation}
G_{+x-polarized} = {{e^2}\over{4 h kT}} \int_0^{\infty} dE |T_{tot} (E)|^2 
sech^2 
\left ( {{E - 
E_F}\over{2 kT}} \right ),
\label{conductance}
\end{equation}
where
\begin{equation}
|T_{tot} (E)|^2 = |T_{1} (E)|^2 + ({k_{x}}^d / {k_{x}}^u ) |T_{2} (E)|^2
\end{equation}

Similarly, the conductance of the minority spin carriers ($G_{-x-polarized}$) is 
calculated after repeating the scattering problem for electrons incident from 
the minority spin 
band in the contacts.  Since the +x and -x-polarized spin states are orthogonal 
in the 
contacts, 
the total conductance of the spin interferometer is  given by
\begin{equation}
G =  
G_{+x-polarized}  + G_{-x-polarized}.
\end{equation}

\subsection{Role of the interface potentials}

The interface potentials $V_I$ determine the solutions of the Schr\"odinger 
equation, and therefore the transmission probabilities and the conductance. To 
elucidate the role of $V_I$, we introduce the following parameter
\begin{equation}
Z = \frac{ 2 {m_f}^* V_0 }{ {\hbar}^2} 
\end{equation}

Typical values of $Z$ vary in the range of 0 to 2 \cite{nitta,schapers2}.
Using ${m_f}^* = m_0$ and $k_F $ = 1.05x$10^8 $ cm$^{-1}$, we get a barrier 
strength $V_0$ = 16 eV-$\AA$ for Z = 2. In the next section, we will show
how the conductance modulation of the spin interferometer depends on $Z$.

\newpage
\section{Numerical Examples:} 
We consider a spin interferometer consisting of 
a quasi one-dimensional InAs channel between two ferromagnetic contacts.  The 
electrostatic potential in the 
z-direction is assumed to be harmonic (with $\hbar \omega$ = 10 meV in Equation 
(4)).
A Zeeman splitting energy of 0.34 meV is used in the semiconductor channel 
assuming a magnetic field B = 1 Tesla along the channel. 
This corresponds to a  $g^*$ factor of 3 and an electron effective mass $m^* = 
0.036 m_o$ which is typical of InAs-based channels \cite{datta}. The Fermi level 
$E_f$ and the exchange splitting energy $\Delta $ in the ferromagnetic contacts 
are set equal to 4.2 and 3.46 eV, respectively  \cite{mireles}.

The Rashba spin-orbit coupling strength ${\alpha}_R$ is 
typically derived from low-temperature magnetoresistance measurements 
(Shubnikov-de Haas oscillations) in 2DEG created at the interface of 
semiconductor heterostructures \cite{nitta}. To date, the largest reported 
experimental values of the Rashba spin-orbit coupling strength ${\alpha}_R$ has 
been found in 
InAs-based semiconductor heterojunctions. For a normal HEMT 
$In_{0.75} Al_{0.25} As/In_{0.75} Ga_{0.25} As $ heterojunction, Sato et al. 
have reported variation 
of ${\alpha}_R$ from 30- to 15 $\times 10^{-12}$ eV-m when the external gate 
voltage is swept from 0 to -6 V (the total electron 
concentration in the 2DEG is found to be reduced from 5- to 4.5$\times 10^{11} 
/cm^2$ over the same range of bias). 
For a channel length of 0.2 $\mu m$, this corresponds to a variation of the
spin precession angle $\theta$ = $2 k_R L$ from about $\pi$ to 0.5$\pi$ over 
the same range 
of gate bias.

In the numerical results below, we calculated the conductance of a spin 
interferometer with a 0.2 $\mu m$ long channel
as a function of the gate voltage at a temperature of 2 K \cite{numerical}.
Tuning the gate voltage varies both  the potential 
energy barrier $\Delta E_c$ and the Rashba spin-orbit coupling strength 
$\alpha_R$. Both of these variations lead to distinct types of conductance 
oscillations. The variation of $\Delta E_c$ causes the Fermi-level in the 
channel to sweep through the resonant energies in the channel, causing the 
conductance 
to oscillate. 
%NEW STUFF
These are known as Ramsauer oscillations (or Fabry-Perot-like resonances)
and have been examined in the past by Matsuyama et al. \cite{add2} for 
two-dimensional structures and by us \cite{cahay_prl} for one-dimensional 
structures. 
%END OF NEW STUFF
The variation of $\alpha_R$, on the other 
hand, causes spin precession in the channel leading to the type of 
conductance oscillation which is the basis of the spin interferometer, as 
originally visualized by Datta and Das \cite{datta}. In
ref. \cite{cahay_prl} we found that the Ramsauer oscillations are much stronger 
and can mask the oscillations due to spin precession, unless the 
structure is designed with particular care to eliminate (or reduce) the Ramsauer 
oscillations.  In the calculations reported here, we vary $\Delta E_c$ over a 
range of 10 meV which allows us to display several of the Ramsauer oscillations 
in the 
conductance. We are restricted to this range because we can increase 
$\Delta E_c$ at most by an amount equal to the Fermi energy in the channel. At 
the end of this range, the Fermi energy lines up with the conduction band edge 
in the channel which 
corresponds to onset of complete pinch-off, i.e., the channel carrier 
concentration falls to zero.
Therefore, the maximum range of $\Delta E_c$ is the Fermi energy, as long as 
we are applying a negative gate voltage to deplete the channel as opposed to
applying a positive gate voltage to accumulate the channel (we do not want to 
accumulate the channel since a large carrier concentration in the channel 
will ultimately shield the gate potential resulting in loss of gate control).
In typical semiconductor channels, the carrier concentration will correspond to 
a Fermi energy of 10 meV, and this dictated our choice for the range of $\Delta 
E_c$.

Over this range of $\Delta E_c$, we assume that the Rashba spin-orbit 
coupling strength ${\alpha}_R$ 
varies from 30 $\times 10^{-12}$ eVm down to zero. This is consistent with 
experimentally observed dependence of ${\alpha}_R$ on gate voltage. This 
variation of ${\alpha}_R$ corresponds to a 
variation of the spin precession angle 
$\theta$ from about $\pi$ to 0 (i.e. half a cycle of the oscillation expected 
from
spin precession).

\subsection{Influence of the interfacial barrier:} The results of the 
conductance modulation are shown in Fig. 3 for different values of the parameter 
$Z$ 
characterizing the strength of the delta barrier at the 
ferromagnet/semiconductor interface (assumed to 
be the same for both contacts). Instead of plotting the conductance as a 
function of gate voltage,
we always plot it as a function of $\Delta E_c$ since $\Delta E_c$ directly 
enters the Hamiltonian in Equation (4). The exact relationship between $\Delta 
E_c$ and the gate voltage are complicated by many factors (interface states, 
channel geometry, etc.), but for the sake of simplicity, we will assume that 
$\Delta E_c$ depends linearly on gate voltage. Therefore, the plots in Figs. 3-9 
can be effectively viewed as plots of conductance versus gate voltage.

A value of Z = 1 corresponds to a value of $ V_L $ and $V_R$ in Equation (2) 
equal to 8 eV-$\AA$.  Figure 3 shows that the location of conductance minima and 
maxima are 
only slightly shifted along the $\Delta E_c $ axis with the variation of the 
parameter Z. The amplitudes of the oscillations increase with Z but then start 
to decrease as the maxima of 
the conductance is reduced for larger values of Z. This reduction in amplitude
is expected since the conductance of the spin interferometer eventually reduces 
to zero as Z $\rightarrow \infty$ (no electron can enter or exit the channel if 
there are infinite barriers at the contact interface).  The maximum in the 
conductance amplitude modulation occurs for Z = 0.25 in our 
numerical examples. In the subsequent numerical simulations which investigate 
the influence of 
impurity scattering on the conductance modulation, we therefore used Z = 0.25 
throughout.

\vskip .2in
\subsection{Impurity scattering:} First, we consider the case of a single 
repulsive 
impurity at a fixed location within the channel ( 300 $\AA$ from the left 
ferromagnetic contact) but with varying strength ${\Gamma}_i$.   
Figure 4 shows that the size and location of the conductance peaks and minima 
are affected by the strength of the impurity scatterer, and more strongly 
affected at larger values of $\Delta E_c $. This is expected since the 
transmission probability through the impurity diminishes as the channel 
approaches pinch-off. Even though not 
shown here, the same trend was observed when the impurity was assumed to be an 
attractive scatterer (negative value for ${\Gamma}_i$).  Figures 5 and 6 
illustrate the dependence of the 
conductance of the interferometer on the exact location of an impurity with a 
scattering strength of ${\Gamma}_i = 0.5 eV\AA$.
Figures 5 and 6 correspond to the case of a repulsive and attractive impurity, 
respectively.
These figures clearly show that the conductance modulation of the interferometer 
operating
in a phase coherent regime is affected by the exact location and strength of a 
single scatterer.

Next, we consider the case of two impurities in the channel at two different 
locations (300 $\AA$,1000 $\AA$) and (500$\AA$,1250 $\AA$). The results for the 
cases of attractive and repulsive impurities (of equal strength) are shown in 
Figures 7 and 8, respectively.
These figures accentuate even more the features observed in Figs. 5 and 6, i.e., 
a strong dependence
of the oscillation amplitude and phase (even far from pinch off)  on the 
impurity type and configurations. 
This sensitivity is due to the quantum interference between electron waves 
reflected multiple times between  impurities and also between each impurity and 
the 
closest ferromagnetic contact.  All these interferences affect the overall 
transmission 
probability of an electron through the  interferometer, and hence its 
conductance. These simulations show that, even if  
good ferromagnetic/semiconductor contacts with large degree of spin polarization 
can be realized through the use of an 
appropriate interfacial barrier, perfect control of the location of the 
conductance minima and maxima could still 
be elusive in the presence of just a few impurities in the channel. Obviously, 
this will have a deleterious 
effect on device reproducibility.

The strong sensitivity to the presence of impurities in the channel also
has a profound influence on the spin-conductance polarization  which is defined 
as 
\begin{equation}
P = \frac{ G_{+x-polarized} - G_{-x-polarized} }{ G_{+x-polarized} + 
G_{-x-polarized} 
}.
\end{equation}
This quantity is plotted in Fig. 9 as a function of $\Delta E_c$. The degree of 
spin polarization $P$ is shown for the case of 
an impurity free channel, and also for the four different two-impurity 
configurations (attractive and repulsive) 
considered in Figures 7 and 8. This quantity
takes both positive and negative values as the gate voltage is swept, and 
reaches a maximum of $60 \%$ close to 
the threshold for channel pinch-off. However, near pinch-off, our model of 
impurity scattering 
should be modified to take into account the absence of screening at low carrier 
density. Even for a more refined model of impurity scattering, we believe that 
Fig. 9 is 
indicative of what is to be expected in realistic samples, i.e, the 
spin-conductance polarization is very 
sensitive to the nature and location of the impurities in the channel. The spin 
polarization therefore provides an 
actual fingerprint for each impurity configuration, a phenomenon similar to the 
universal conductance fluctuations 
linked to the displacement of a single impurity in mesoscopic samples 
\cite{stone}.

\newpage
\vskip .2in
\section{Conclusions:} 
In this paper, we have developed a fully quantum mechanical approach to 
model coherent electron spin transport in a disordered semiconductor channel 
using a 
particular model of impurity scattering. 
We have also shown 
how conductance modulation of gate controlled spin interferometers proposed in 
ref. \cite{datta}  are affected by the presence of interfacial
barriers at the ferromagnetic contact/semiconductor interfaces and 
also by a few impurities in the semiconducting channel. 
Quantum interference caused by multiple reflections of electron waves between 
impurities, 
and between the impurities and the interfacial barriers, can strongly affect the 
overall degree of spin 
polarization of the interferometer. The extreme sensitivity of the amplitude and 
phase of conductance oscillations to  impurity
location is reminiscent of the phenomenon of universal conductance fluctuations
of mesoscopic samples. This will hinder practical 
applications of electron spin interferometers.

\vskip .1in
The work of  S. B. is supported by the
National Science Foundation under grant ECS-0089893.

\pagebreak

\begin{center}
{\bf Figure Captions}
\end{center}

\vskip .3in
\noindent {\bf Fig. 1}: A schematic of the electron spin interferometer
from ref. \cite{datta}.  The horizontal dashed line represents the quasi 
one-dimensional electron gas formed at the semiconductor interface between 
materials I and II.  The magnetization of the ferromagnetic contacts is assumed 
to be along the +x-direction which results in a magnetic field along the 
x-direction.  
Also shown is a qualitative representation of the energy dispersion of the two 
perturbed (solid line) and unperturbed (broken line)  bands under the gate - the 
perturbation is due to the axial magnetic field along the channel.  

\vskip .2in
\noindent {\bf Fig. 2}: Energy band diagram across the electron spin 
interferometer.  We use a Stoner-Wohlfarth model for the ferromagnetic contacts. 
$\Delta$ is the exchange splitting energy in the contacts. 
$\Delta E_c$ is the height of the potential barrier between the energy band 
bottoms of 
the semiconductor and the ferromagnetic electrodes. $\Delta E_c$ takes into 
account 
the effects of the quantum confinement in the y- and z-directions.  Also shown 
as dashed lines are the resonant energy states above $\Delta E_c$. Peaks in the 
conductance of the electron spin interferometer are expected when the Fermi 
level in the contacts lines up with the resonant states. The barriers at the 
ferromagnet/semiconductor interface are modeled as simple one-dimensional 
delta-potentials.

\vskip .2in
\noindent {\bf Fig. 3}: 
Conductance modulation of a ballistic electron spin
interferometer  (for T = 2 K) as the gate voltage (or the energy barrier $\Delta 
E_c$) is varied.  We assume that the Rashba coupling strength ${\alpha}_R$ 
varies 
from 30 $\times 10^{-12}$ eVm to 0 for the range of $\Delta E_c$ shown on the 
figure. This should correspond to one-half cycle of conductance oscillation due 
to spin precession.  
The separation between the two ferromagnetic contacts is 0.2 $\mu m$.  
The confinement energy $\hbar \omega$ along the z-direction (direction 
transverse to both current flow and the gate electric field) is 10 meV. 
The conductance oscillations in this figure are caused by the Fermi level 
sweeping through the resonant levels in the channel of the interferometer 
(the so-called Ramsauer effect) and are {\it not} due to the spin precession in 
the channel as shown in ref. \cite{cahay_prl}. The different curves correspond 
to 
different values of the parameter $Z$ characterizing the strength of the 
interfacial barrier between the ferromagnetic contact and semiconducting 
channel. 
The semiconducting channel is assumed to be impurity free, and hence ballistic.

\vskip .2in
\noindent {\bf Fig. 4}: Influence of a single impurity on the conductance 
modulation of an electron spin interferometer. All other parameters are the same 
as in Fig.3. The interface potential
at the ferromagnet/semiconductor interface is 2 eV-$\AA$ corresponding to  $Z$ = 
0.25. The impurity is modeled as a repulsive 
delta-scatterer with strength ${\Gamma}_i$ indicated next to each curve 
in unit eV-$\AA$.  The impurity is located 
300 $\AA$ away from the left ferromagnetic contact/channel interface.

\vskip .2in
\noindent {\bf Fig. 5}: Influence of a single impurity on the conductance 
modulation of an electron spin interferometer. Again, all other parameters are 
the same as in Fig.3, and $Z$ = 0.25. The impurity is modeled as a repulsive 
delta-scatterer with strength $\Gamma$ = 0.5 eV-$\AA$.  
Cases 1 through 4 correspond to an impurity located 300, 750, 1000, 
and 1500 $\AA$ away from the left ferromagnetic contact/channel interface.

\vskip .2in
\noindent {\bf Fig. 6}: 
Same as Figure 5 for the case of an attractive impurity with strength
$\Gamma$ = - 0.5 eV$\AA$. Cases 1 through 4 correspond to an impurity 
located 300, 750, 1000, and 1500 $\AA$ away from the left 
ferromagnetic contact/channel interface.
 
\newpage

\vskip .2in
\noindent {\bf Fig. 7}: 
Same as Figure 5 for the case of two repulsive impurities with strength
$\Gamma$ = 0.5 eV$\AA$. The curves labeled 1 and 2 correspond to the case of
two impurities located at (300 $\AA$,1000 $\AA$) and (500$\AA$,1250 $\AA$), from 
the left ferromagnet/channel interface, respectively.

\vskip .2in
\noindent {\bf Fig. 8}: 
Same as Figure 5 for the case of two attractive impurities with strength
$\Gamma$ = - 0.5 eV$\AA$. The curves labeled 1 and 2 correspond to the case of
two impurities located at (300 $\AA$,1000 $\AA$) and (500$\AA$,1250 $\AA$), from 
the left ferromagnet/channel interface, respectively.

\vskip .2in
\noindent {\bf Fig. 9}: 
Degree of spin-conductance polarization P versus $\Delta E_c$. All other 
parameters are the same as listed in Fig.3. The quantity $P$
is plotted for the case of a ballistic channel with no impurity,
and also for the four two-impurity configurations (attractive and repulsive) 
considered in Figures 7 and 8. The curves labeled 1 and 2 correspond to the case 
of
two impurities located at (300 $\AA$,1000 $\AA$) and (500$\AA$,1250 $\AA$), from
the left ferromagnet/channel interface, respectively. The extra labels ``r'' and 
``a''
are to identify the case of repulsive and attractive scatterers, respectively.

\newpage
\
%Figure 1
\vskip 1in
\begin{figure}[h]
\centerline{\psfig{figure=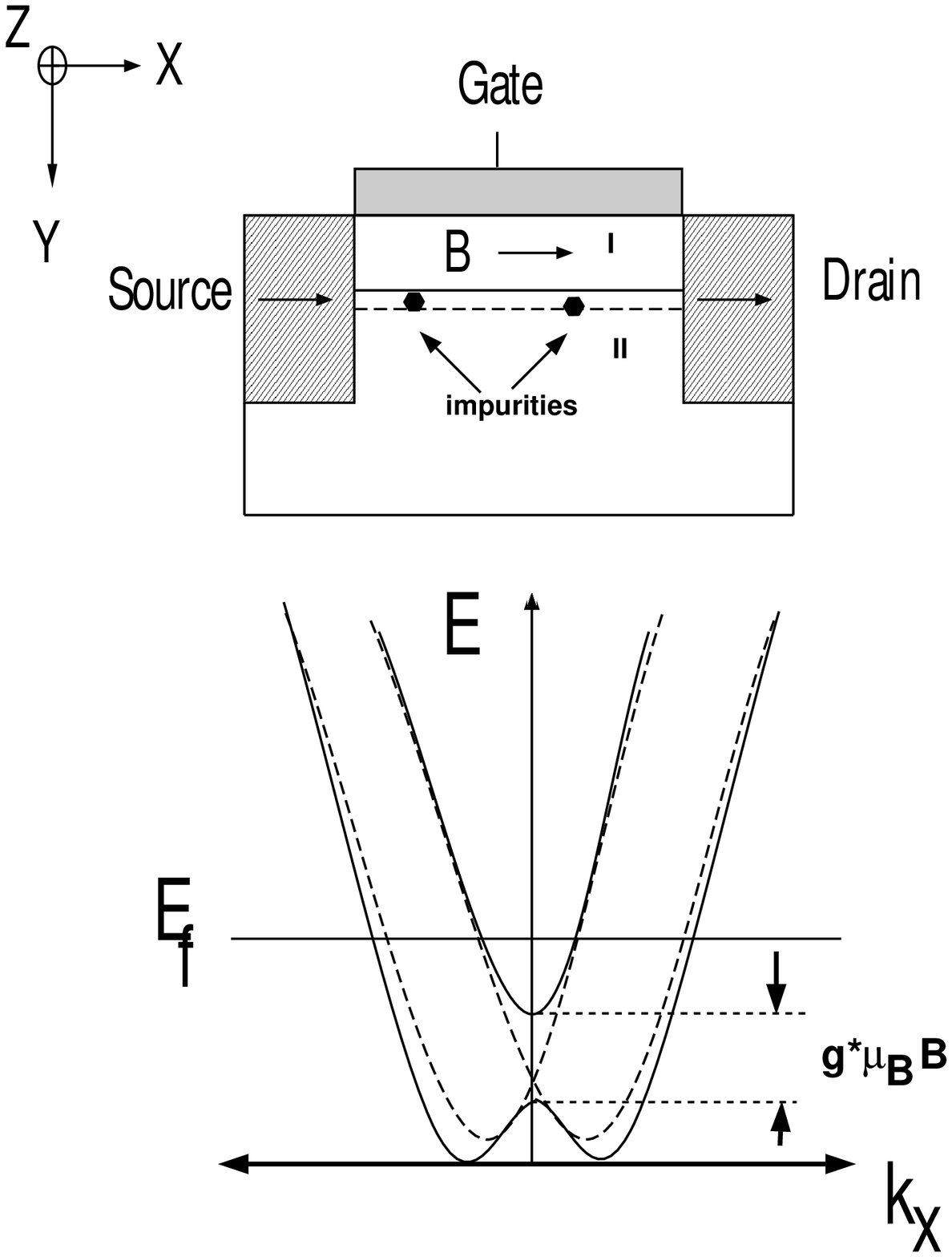,height=6in,width=6in}}
\end{figure}

\newpage
\
%Figure 2
\vskip .1in
\begin{figure}[h]
\centerline{\psfig{figure=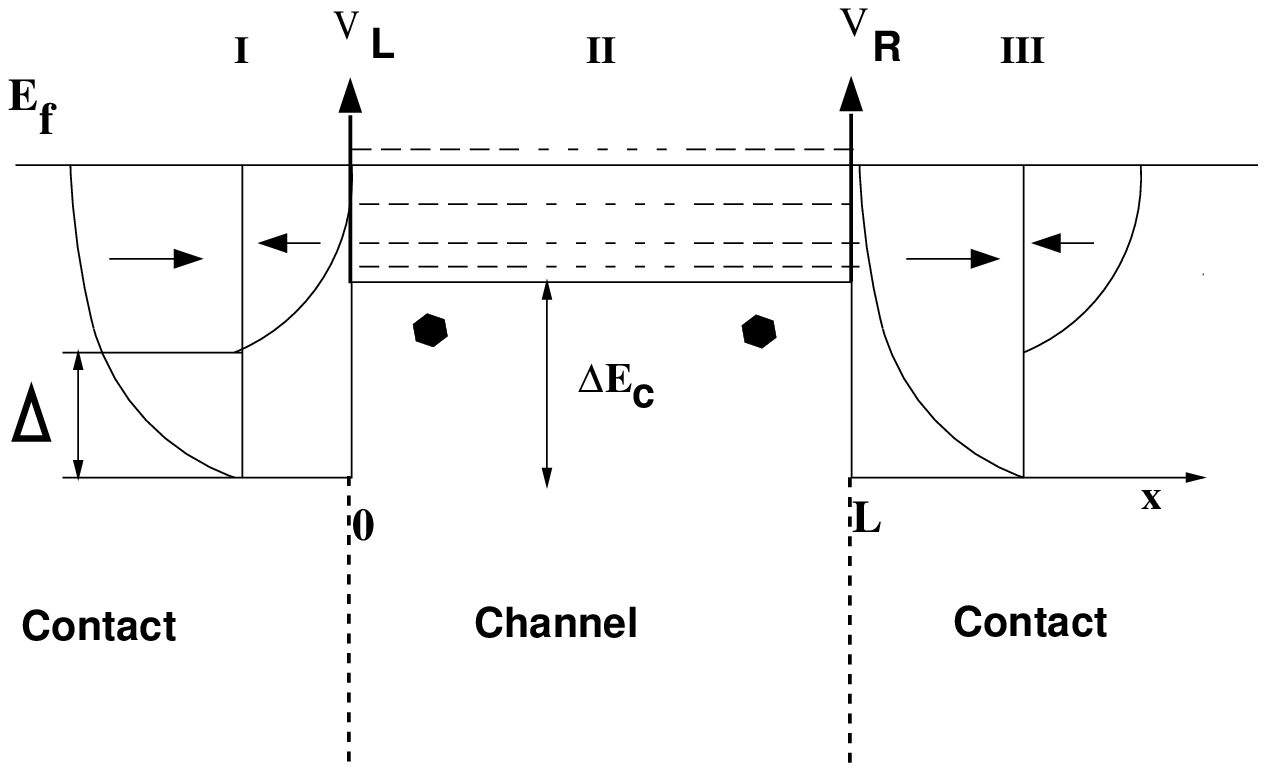,height=7in,width=6.5in}}
\end{figure}

\newpage
\
%Figure 3
\vskip .3in
\begin{figure}[h]
\centerline{\psfig{figure=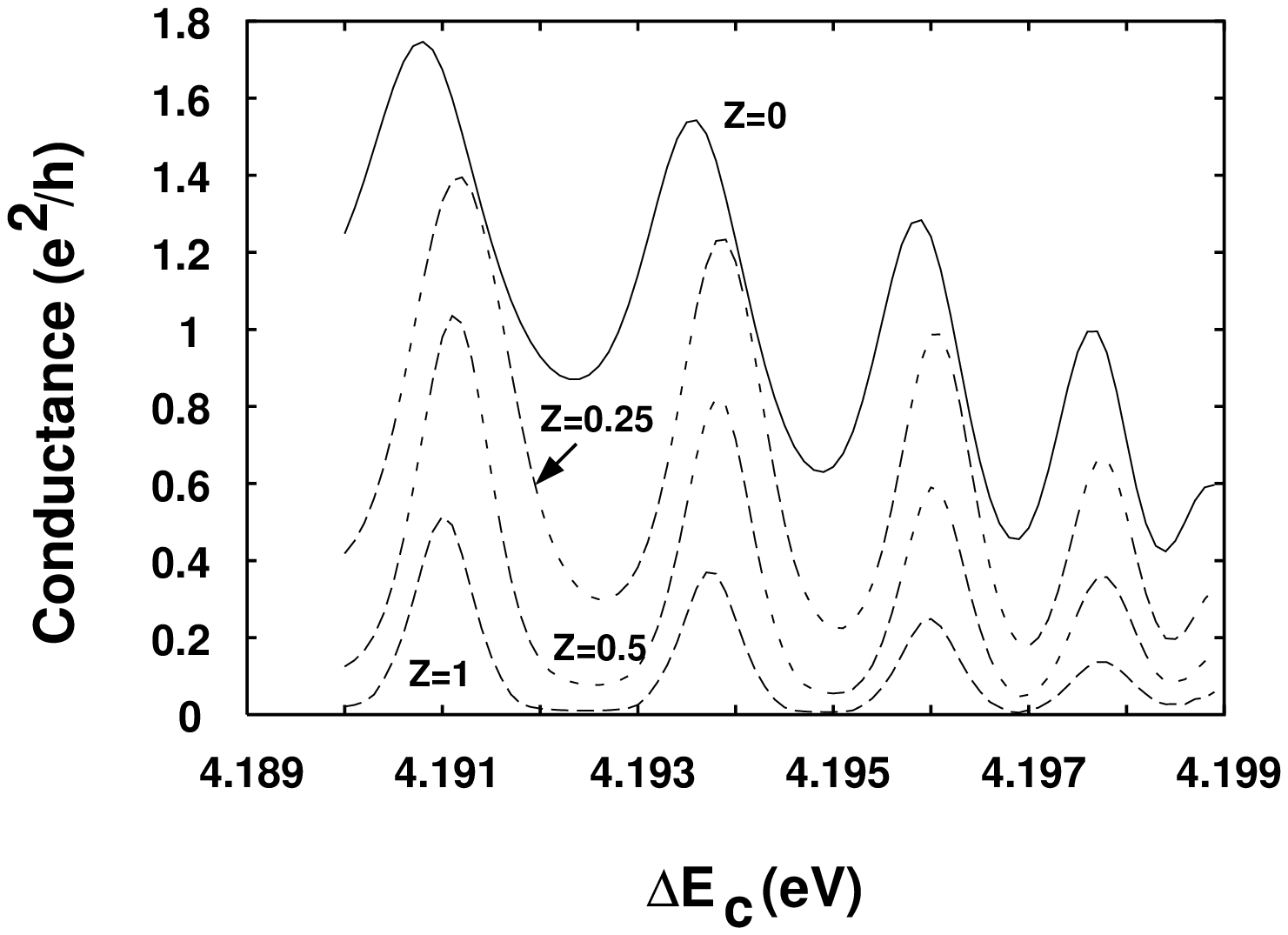,height=7in,width=7in}}
\end{figure}

\newpage
\
%Figure 4
\vskip 1in
\begin{figure}[h]
\centerline{\psfig{figure=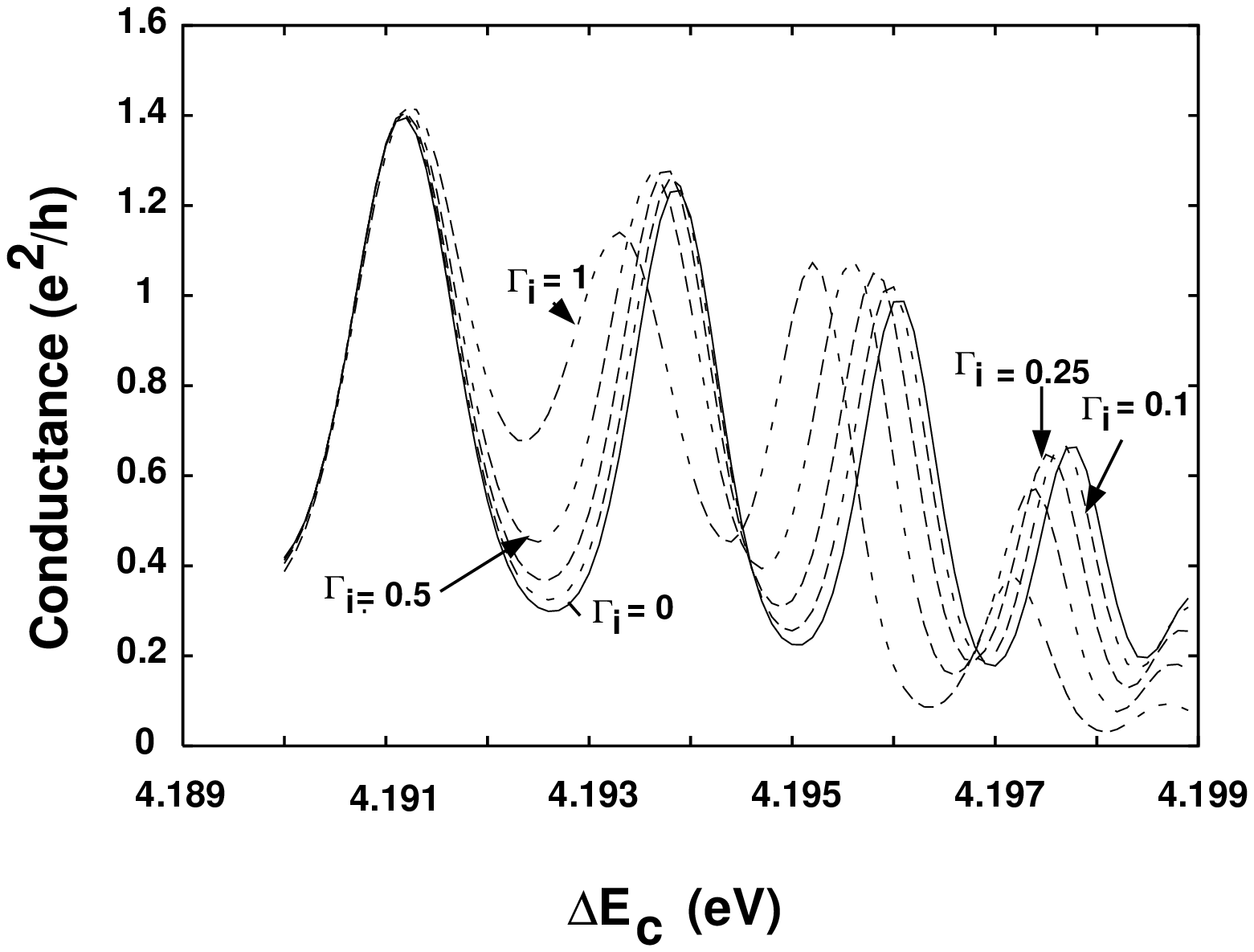,height=7in,width=7in}}
\end{figure}

\newpage
\
%Figure 5
\vskip 1in
\begin{figure}[h]
\centerline{\psfig{figure=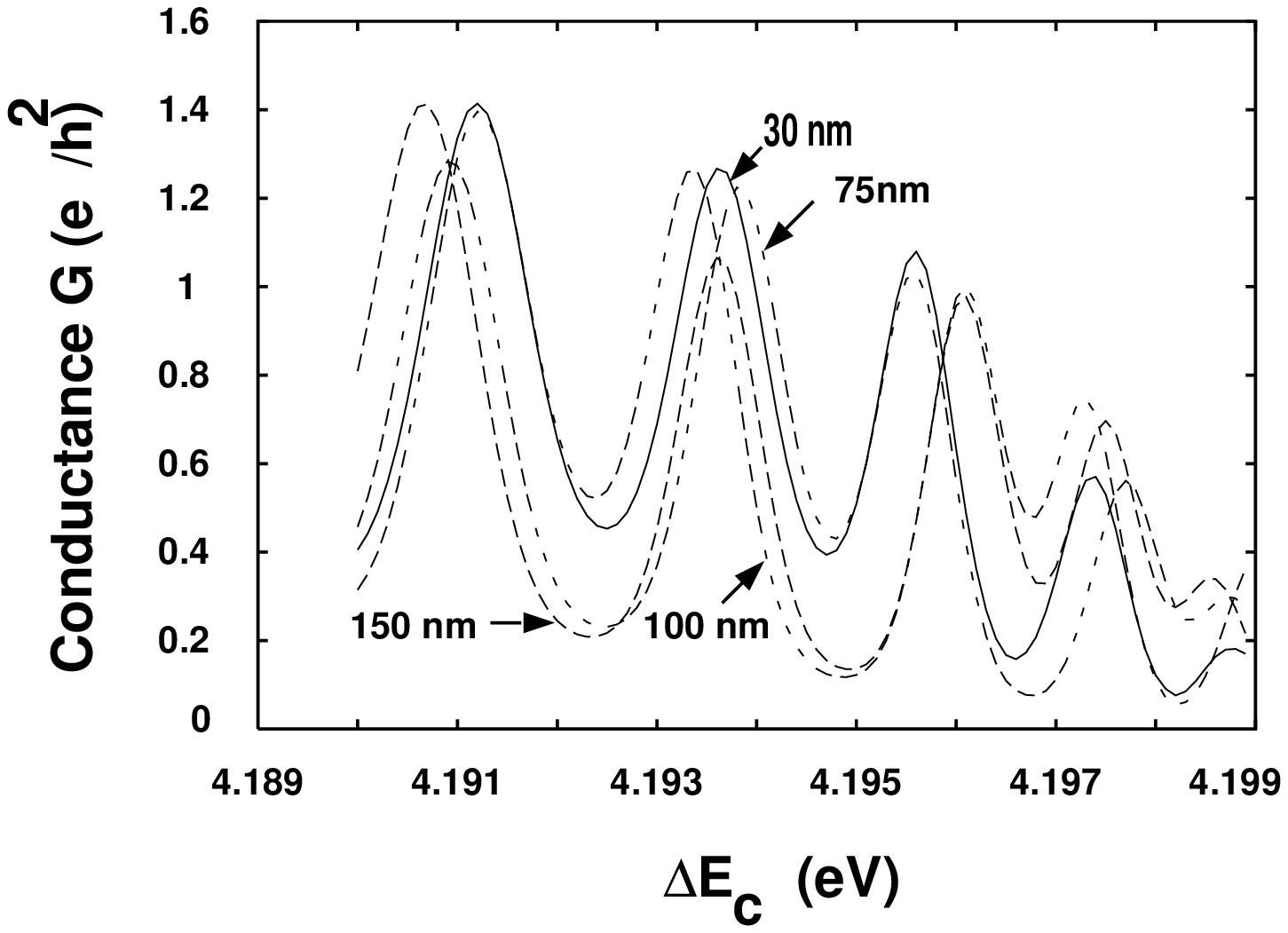,height=7in,width=7in}}
\end{figure}

\newpage
\
%Figure 6
\vskip 1in
\begin{figure}[h]
\centerline{\psfig{figure=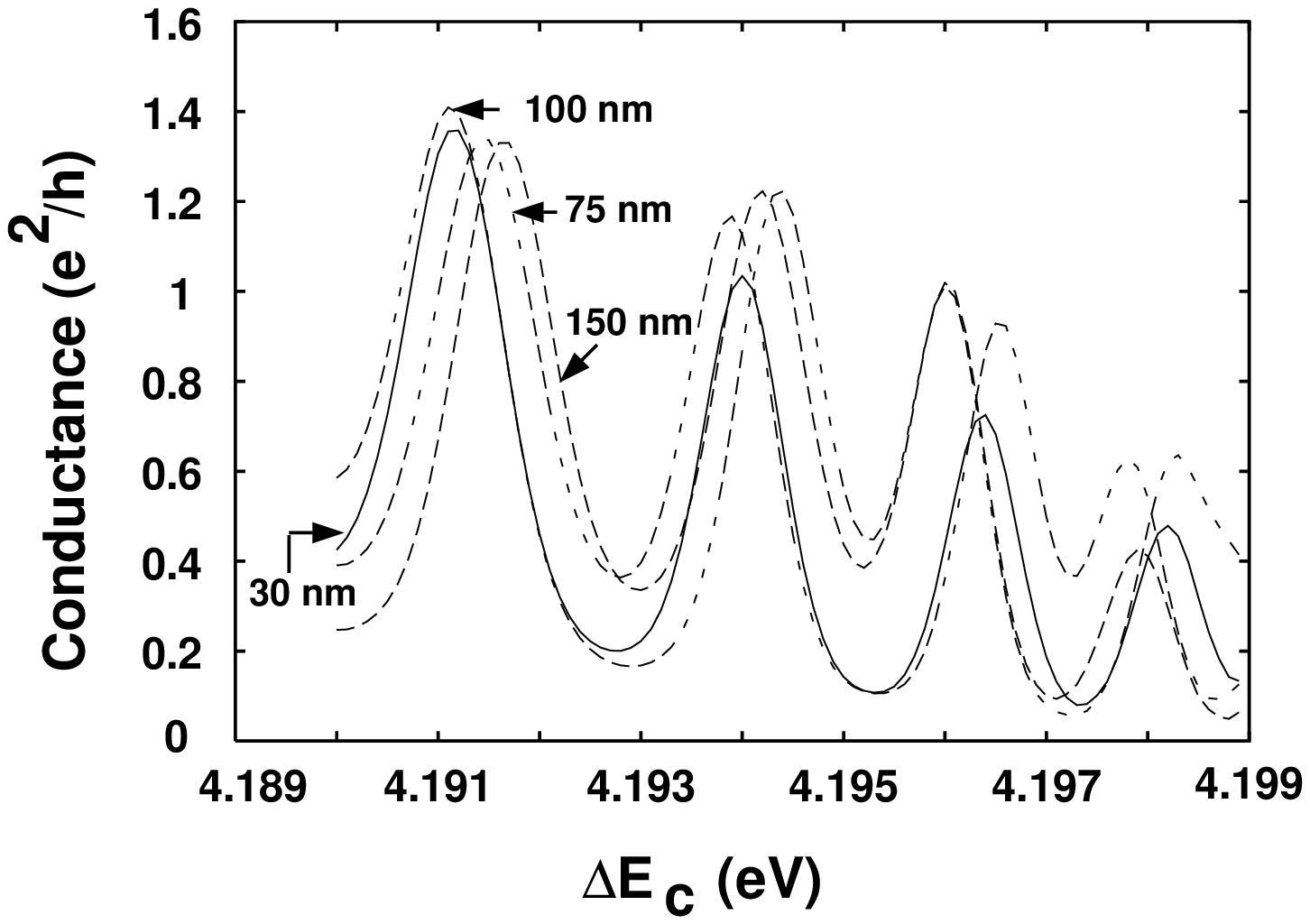,height=7in,width=7in}}
\end{figure}

\newpage
\
%Figure 7
\vskip 1in
\begin{figure}[h]
\centerline{\psfig{figure=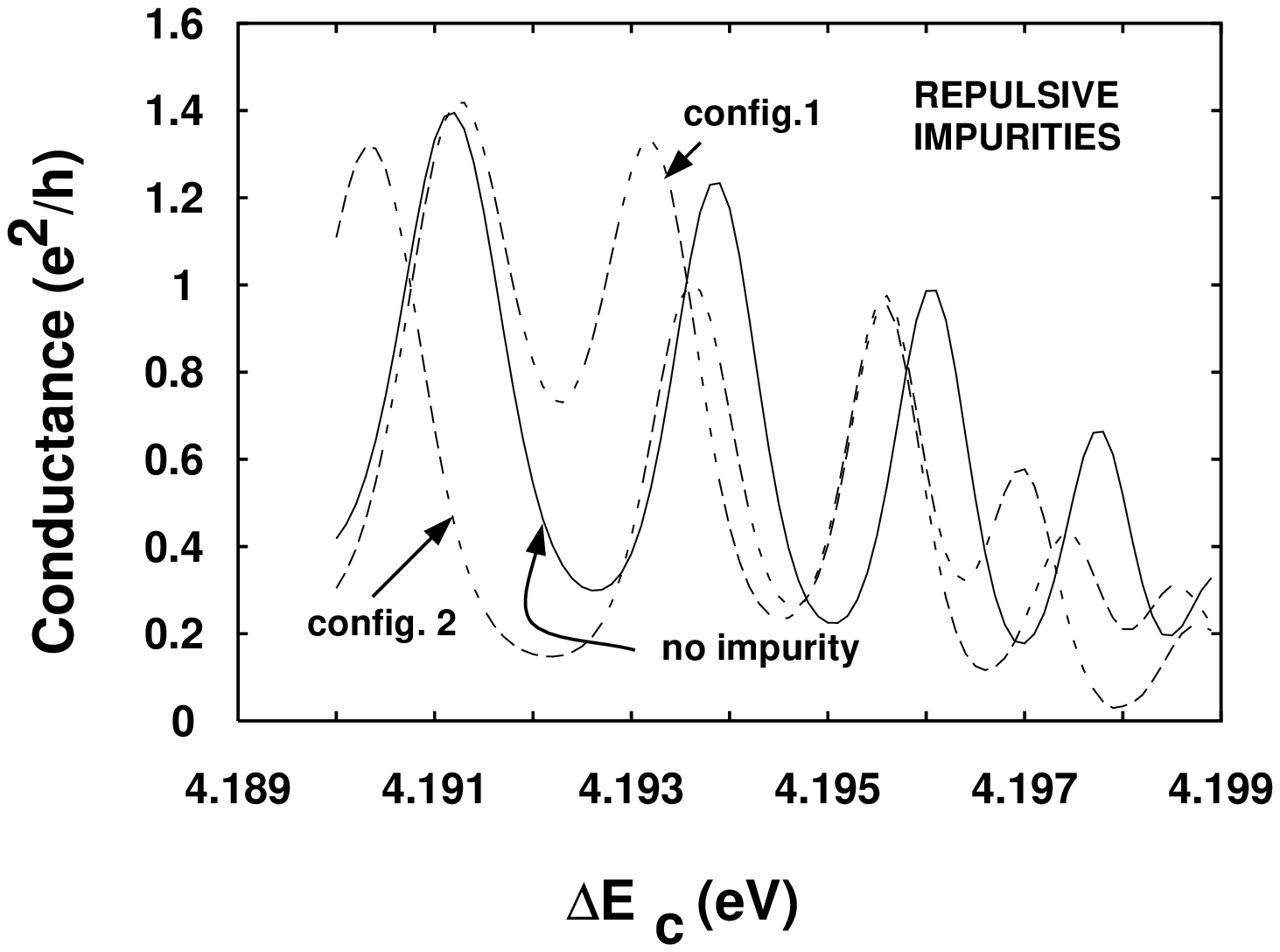,height=7in,width=7in}}
\end{figure}

\newpage
\
%Figure 8
\vskip 1in
\begin{figure}[h]
\centerline{\psfig{figure=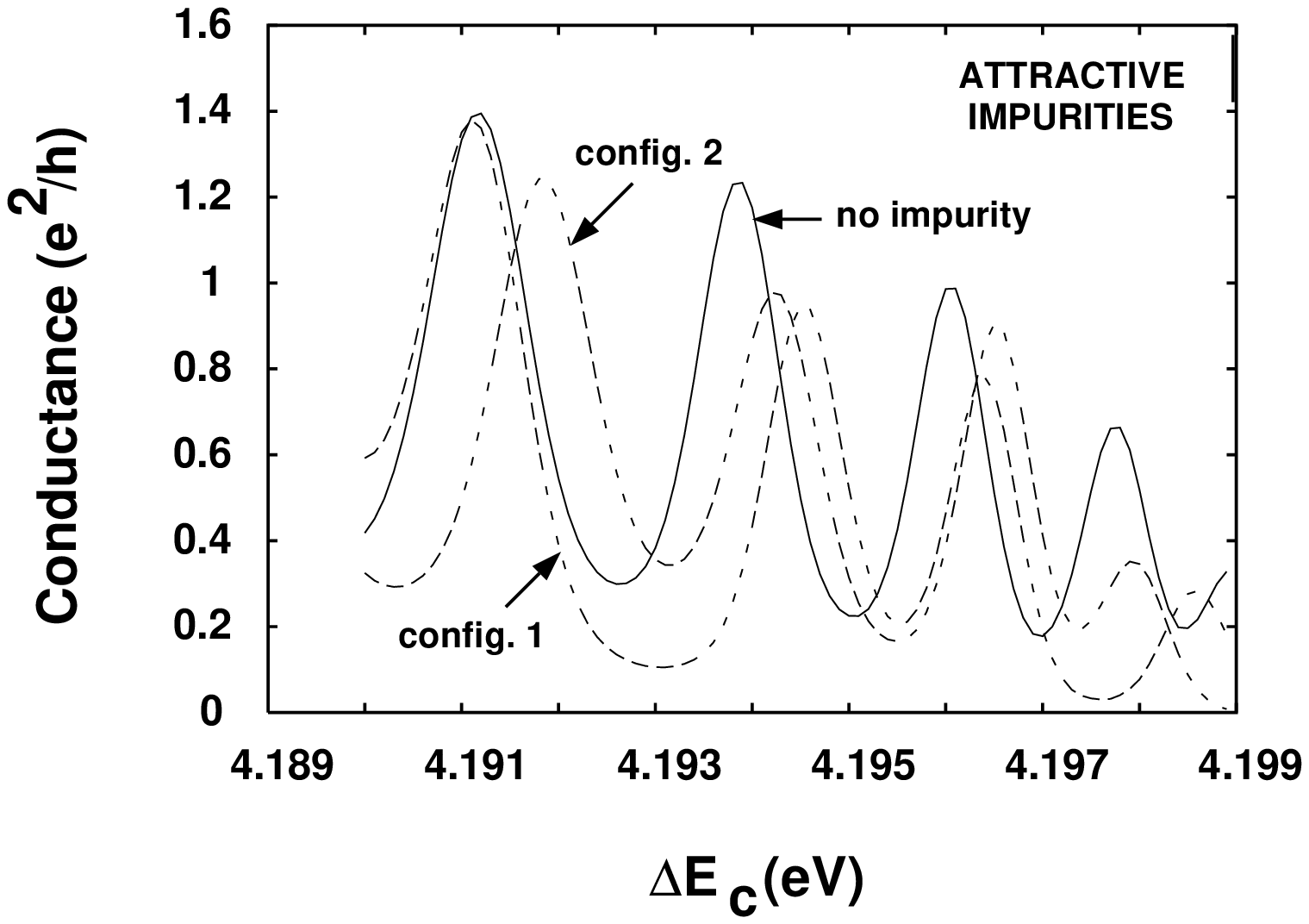,height=7in,width=7in}}
\end{figure}

\newpage
\
%Figure 9
\vskip 1in
\begin{figure}[h]
\centerline{\psfig{figure=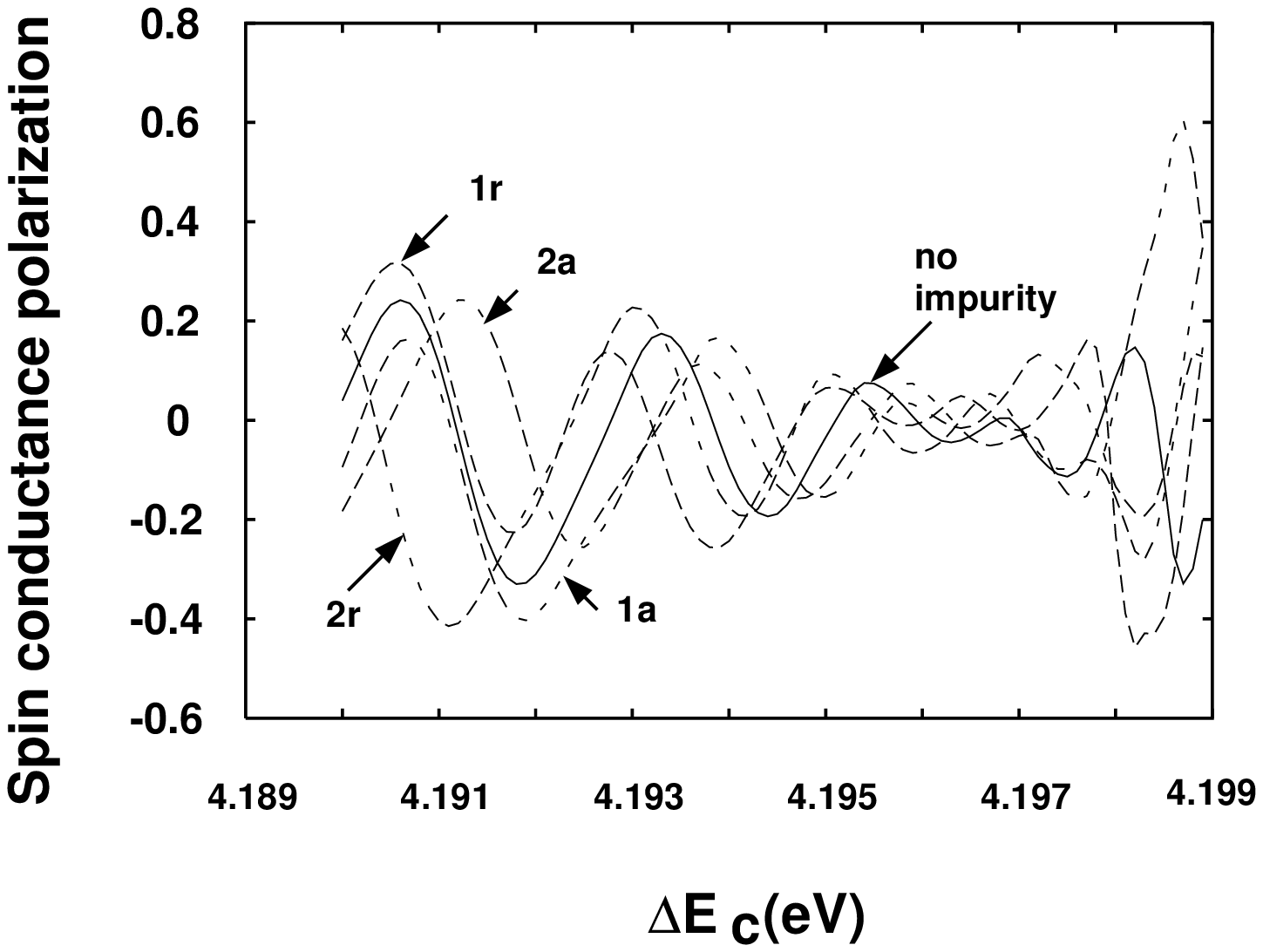,height=7in,width=7in}}
\end{figure}


\begin{thebibliography}{10}
\bibitem{numerical}
We compute the conductance using the Landauer formula by integrating
over an energy range from [ $E_f - 4 k_B T, E_f + 4 k_B T $]. For each 
temperature, we limit the range of variation of $\Delta E_c$ so that both 
channels under the 
gate are conducting for the range of energy considered.
\end{thebibliography}
\end{document}